\PassOptionsToPackage{dvipsnames}{xcolor}
\documentclass[sigconf]{acmart}

\AtBeginDocument{%
  }

\copyrightyear{2026}
\acmYear{2026}
\setcopyright{cc}
\setcctype{by}
\acmConference[HASP '26]{International Workshop on Hardware and Architectural Support for Security and Privacy 2026}{October 31-November 01, 2026}{Athens, Greece}
\acmBooktitle{International Workshop on Hardware and Architectural Support for Security and Privacy 2026 (HASP '26), October 31-November 01, 2026, Athens, Greece}
\acmDOI{10.1145/3857307.3857312}
\acmISBN{979-8-4007-2964-5/2026/10}

\usepackage{booktabs}
\usepackage{balance}
\usepackage{xspace}
\usepackage{xcolor}

\usepackage{amsmath}
\usepackage{algorithmic}
\usepackage{graphicx}
\usepackage[dvipsnames]{xcolor}

\usepackage{float}   

\usepackage{booktabs}
\usepackage{xspace}
\usepackage{enumitem}
\usepackage{hyperref}
\usepackage[htt]{hyphenat}
\usepackage{tikz}
\usetikzlibrary{arrows.meta,decorations.pathreplacing}
\usetikzlibrary{calc}

\newcommand{\tool}{\textsc{zkDuel}\xspace}

\begin{document}

\title{zkDuel: Characterizing SumCheck Prover Kernels across GPU Programming Models}

\author{Samarth Narang}
\affiliation{%
  \department{Tandon School of Engineering}
  \institution{New York University}
  \city{Brooklyn}
  \state{NY}
  \country{USA}
}
\email{sn2910@nyu.edu}

\author{Alhad Daftardar}
\affiliation{%
  \department{Tandon School of Engineering}
  \institution{New York University}
  \city{Brooklyn}
  \state{NY}
  \country{USA}
}
\email{ajd9396@nyu.edu}

\author{Brandon Reagen}
\affiliation{%
  \department{Tandon School of Engineering}
  \institution{New York University}
  \city{Brooklyn}
  \state{NY}
  \country{USA}
}
\email{bjr5@nyu.edu}

\renewcommand{\shortauthors}{Narang, Daftardar, and Reagen}

\begin{abstract}
Zero-knowledge proofs have attracted growing interest from the hardware
acceleration community, driven by their use in verifiable computation,
privacy-preserving systems, and blockchain applications. A key prover kernel in
many such systems is SumCheck, whose table-parallel structure makes it a
natural target for GPU acceleration. Practical SumCheck kernels over
cryptographic fields, however, are shaped not by bandwidth but by multi-limb
arithmetic, Montgomery reduction, expression specialization, reduction
structure, and launch overhead.

We present \tool, an open-source benchmark that isolates SumCheck as a
standalone GPU kernel through matched Triton and CUDA implementations.
\tool{} pairs diagnostic microbenchmark polynomials for controlled analysis
with zkPHIRE-derived constraints for prover-style stress testing, sweeps
32/64/128/256-bit prime fields, and validates both backends against a CPU
reference before any performance measurement.

Across 800 attempted configurations on an NVIDIA H100, 758 complete with
matching Triton and CUDA checksums, and both backends reproduce the CPU
reference on every reported workload and field width. On these configurations, we find that the cost of
programming-model abstraction reduces with arithmetic width. CUDA holds a
$3.5\times$ median advantage over Triton at 32 bits but only $1.4\times$ at
256 bits, where 30 of 171 validated configurations finish within $1.1\times$ of
CUDA, including 19 large-table configurations on which Triton is faster. An
A100 exhibits the same width-dependent convergence, from $4.4\times$ at 32 bits
to $1.6\times$ at 256 bits. This
convergence comes from fixed overheads being amortized as
per-element arithmetic grows. In particular, host-side launch costs dominate
short-running Triton kernels, while Triton's original block-reduction
implementation performs redundant modular additions across threads. Replacing
that reduction with \texttt{tl.reduce} brings Triton's median kernel time to
within 8\% of CUDA's on both A100 and H100. Specialized point interpolation
accelerates 256-bit evaluation kernels by $1.55$--$1.75\times$ on both
backends.

At large cryptographic field widths, a Python-embedded DSL is therefore a compelling
choice for prover kernels. The remaining performance gap depends strongly on
reduction and launch structure, while arithmetic specialization benefits both
programming models.

\end{abstract}



\maketitle

\section{Introduction}
\label{sec:introduction}

Zero-knowledge proofs have moved from a theoretical construction into deployed
infrastructure, underpinning verifiable computation, privacy-preserving
systems, and blockchain applications. In every such deployment the prover is
the bottleneck: it operates on a secret witness and runs orders of magnitude
slower than verification. SumCheck is a core prover primitive in polynomial proof systems
and plays a central role in Spartan- and HyperPlonk-style constructions~\cite{spartan,hyperplonk2022}.
Its inner loop is highly regular: evaluate a polynomial over paired table entries, reduce
partial sums, derive the next challenge, and fold the table. This structure
makes GPUs attractive. A practical GPU implementation, however, is not a simple
table scan. For 256-bit fields, every logical field operation expands into limb
arithmetic, carry propagation, Montgomery reduction, and temporary storage.
These engineering details determine whether a kernel behaves like a
memory-bandwidth benchmark, an integer-arithmetic benchmark, or a compiler and
code-generation stress test.

Triton has been widely adopted in the machine-learning community as a
Python-integrated DSL for writing custom GPU kernels~\cite{triton}. Its appeal
comes from making high-performance GPU programming more accessible than writing
low-level CUDA \cite{nvidiaCUDA} directly, especially for tensor-oriented floating-point
workloads such as tiled matrix operations, reductions, and fused elementwise
kernels. Cryptographic prover kernels stress different mechanisms: multi-limb
integer arithmetic, modular reduction, carry propagation, and layout-sensitive
table operations. The question is therefore not whether such a programming
model can express finite-field prover kernels, which it can, but what the
abstraction costs, and whether that cost stays constant across the arithmetic
widths cryptography spans.

Answering it requires a native CUDA baseline, because only a matched
implementation separates algorithmic cost from backend cost. Several lines of
work bear on this. Production libraries expose CUDA SumCheck
backends~\cite{icicle}, and recent systems accelerate SumCheck within a
particular prover by partitioning rounds between GPU and host~\cite{chimera}.
Complementary work compares Number Theoretic Transform (NTT)- and SumCheck-based primitives under unified
architectural budgets~\cite{whenproofs}, and a recent performance study of ZKPs on GPUs reports that prover arithmetic
executes on the 32-bit integer pipeline with limited instruction-level
parallelism, leaving it bounded by available integer compute; that study covers
multi-scalar multiplication (MSM) and NTT~\cite{zkprophet}. These efforts target
end-to-end prover performance, other kernels, or platform-independent
comparison. The question we address is narrower, and to our knowledge, remains
open: how does a single prover kernel behave across arithmetic widths and
programming models when the proof system, the workload, and the correctness
criterion are all held fixed?

\tool{} closes that gap. It pairs a CUDA implementation with a
matched Triton implementation of the same 25 prover constraints, derived from
the zkPHIRE constraint set~\cite{zkphire2025}: Spartan-style satisfiability constraints,
Halo2 curve and point-addition, HyperPlonk zero- and
permutation-checks, their Jellyfish variants, and polynomial openings. These constraints
differ in the properties that stress a GPU: product-chain length, number of
distinct input tables, polynomial degree, and count of simultaneously live
intermediates. That variation is what makes them a benchmark rather than one
kernel measured repeatedly. Both backends evaluate identical workload specifications, layouts, and
interpolation settings over four Montgomery prime fields: a 32-bit prime,
Goldilocks, the Mersenne prime $2^{127}-1$, and the BLS12-381 scalar field.

Our central finding is that the cost of programming-model abstraction is not a
fixed tax but falls steadily as arithmetic widens. On an H100, CUDA holds a $3.50\times$ median advantage over Triton at 32 bits;
at 256 bits the gap falls to $1.37\times$, with 30 of 171 validated
configurations finishing within $1.1\times$ of CUDA, including 19 Triton wins.
An A100 exhibits the same width-dependent convergence. We
trace this width dependence to fixed costs that dominate when field arithmetic
is cheap and are progressively amortized as multi-limb Montgomery work grows.
Two effects are particularly important: host-side launch overhead, and
Triton's original block-reduction implementation, in which every thread
performs all of the block's modular additions.
Section~\ref{sec:eval:registers} develops this. Two readings follow: Triton is a plausible implementation vehicle for prover
kernels at the widths cryptography actually uses, and its Python embedding puts
GPU kernel development within reach of implementers who would not write CUDA by
hand, which matters because constraint sets evolve faster than hand-written
kernels can be rewritten. Less expectedly, the result identifies finite-field
cryptography as an application domain for a DSL designed around tiled
floating-point tensors, a workload class with almost none of the properties
Triton was built to exploit. This paper contributes:
\begin{itemize}
  \item \textbf{A characterization of the CUDA--Triton performance gap}
    showing that fixed launch and reduction overheads account for much of the
    gap at narrow fields and are progressively amortized as multi-limb
    arithmetic widens. Replacing Triton's original reduction with
    \texttt{tl.reduce} reduces the median kernel gap to within 8\% of CUDA on
    both A100 and H100.

  \item \textbf{The finding that programming-model overhead decreases with
    arithmetic width}, measured over 758 validated configurations from an
    800-configuration sweep on an H100, with the same trend observed on an
    A100: CUDA's median speedup falls from $3.5\times$ at 32 bits to
    $1.4\times$ at 256 bits, where 30 of 171 configurations finish within
    $1.1\times$ of CUDA and Triton is faster on 19.

  \item \textbf{Backend-independent kernel optimizations for wide-field SumCheck}:
    specialized point interpolation replaces small-point Montgomery
    multiplications with cascading modular additions and speeds up 256-bit
    evaluation kernels by $1.65$--$1.75\times$ in CUDA and
    $1.55$--$1.59\times$ in Triton. A limb-major table layout provides a
    further 1--5\% improvement at 256 bits when combined with specialized
    interpolation.

  \item \textbf{An open, correctness-gated benchmark artifact} with matched
    CUDA and Triton implementations of 25 prover constraints across four
    arithmetic widths, available at
    \url{https://github.com/snarang181/zk-duel}. Reported configurations require cross-backend checksum
    agreement, with both implementations additionally checked against a CPU
    big-integer reference.
\end{itemize}
\section{Background}
\label{sec:background}

\subsection{SumCheck as a Prover Kernel}
\label{sec:bg:sumcheck}

In ZKPs, computations being proven are represented using polynomial constraints
that encode the correctness of a program~\cite{spartan,hyrax2018, hyperplonk2022}, a process called \emph{arithmetization}. The scheme populates
constituent polynomials so that their evaluations at step $i$ describe the
operation performed at that step, and a composite constraint polynomial
vanishes across the execution domain exactly when every step executed
correctly~\cite{plonk2019}. In the implementation, evaluations of the
constituent polynomials are represented as input tables. The prover therefore
needs a way to show that a polynomial vanishes over the entire execution domain
without the verifier inspecting every step directly.

SumCheck provides one mechanism for this global check and appears in several
proof-system constructions~\cite{sumcheck,thaler2013gkr,spartan}. It
lets a prover $\mathcal{P}$ convince a verifier $\mathcal{V}$ that the sum of a
multivariate polynomial $f(X_1,\ldots,X_\mu)$ over the Boolean hypercube equals
a claimed value $C$:
\begin{equation}
  C = \sum_{x_1,\ldots,x_\mu \in \{0,1\}} f(x_1,\ldots,x_\mu).
\end{equation}
The protocol runs for $\mu$ rounds, one per variable $x_i$. In each round the prover
reduces the multivariate polynomial to a univariate polynomial whose degree is bounded by the polynomial’s individual degree in the current variable $x_i$. The verifier checks that evaluating $x_i$ at $0$ and $1$ and computing their sum is
consistent with the previous claim, then samples a random challenge fixing $x_i$ for the next round. After all variables are fixed the verifier
evaluates $f$ once at a random point.

From an implementation standpoint, the prover repeatedly evaluates a round
polynomial over a table of field elements, derives a challenge, and folds the
table for the next round. If the initial table has size $N = 2^r$, each round
halves the active table. Each round views the table as even and odd pairs, and
for a pair $(\mathsf{even}, \mathsf{odd})$ interpolates at a point $p$ as
\begin{equation}
  \label{eq:interp}
  \mathsf{interp}(\mathsf{even},\mathsf{odd},p)
    = \mathsf{even} + (\mathsf{odd}-\mathsf{even}) \cdot p .
\end{equation}
A round evaluates Equation~\ref{eq:interp} at $D+1$ points for every
constituent input table the workload references, where $D$ is the degree of the
round polynomial. Both the table count and $D$ are properties of the constraint
being proven, which is why constraints stress a GPU differently.

This structure makes SumCheck attractive for GPU execution: each round exposes
parallel work over many table entries. The kernel is not a simple dense tensor
operation, however. Each table entry is a finite-field element, and
Equation~\ref{eq:interp} is mathematically trivial but expands, at
cryptographic widths, into multi-limb arithmetic.

\subsection{Prime-Field and Montgomery Arithmetic}
\label{sec:bg:field}

The workloads in \tool{} operate over prime fields $\mathbb{F}_p$, where the
modulus $p$ fixes both the field size $|\mathbb{F}| = p$ and the number of
limbs needed to represent an element. For native-width fields a
field element occupies one machine word; for cryptographic widths each element
occupies several limbs, and \tool{}'s 256-bit path uses eight 32-bit limbs per
element in Montgomery form. Montgomery representation makes repeated modular
multiplication more efficient by replacing explicit division with multiplications~\cite{montgomery1985modular}. Even so, multiplication remains much
heavier than addition: a 256-bit product requires 64 limb-level multiply-add
operations, carry propagation across the partial products, and a reduction
pass. Addition and subtraction are limb-wise carry and borrow chains followed
by a conditional modular correction.

A SumCheck round compounds these costs along two axes. A single application of
Equation~\ref{eq:interp} at 256 bits costs a modular subtraction, a Montgomery
multiplication, and a modular addition, and a round applies it
\texttt{NUM\_INPUTS} $\times (D+1)$ times per pair of entries. Those intermediates
are also simultaneously live, resulting in pressure on registers, instruction
scheduling, and generated-code quality.

This is why 256-bit SumCheck is not simply a scaled-up 32-bit benchmark. As
field width grows, multi-limb arithmetic increasingly shifts the workload away
from memory bandwidth alone and toward arithmetic, reduction, register
allocation, memory layout, and code-generation costs. Product-chain structure
further amplifies these effects by increasing the amount and lifetime of
intermediate state.

\subsection{Arithmetic Width and Soundness}
\label{sec:bg:soundness}
\tool{} sweeps prime fields at four arithmetic widths: 32, 64, 128, and 256
bits. Small fields are attractive because they make prover arithmetic cheaper,
but a single challenge from a 32- or 64-bit field is too small to provide
conventional cryptographic soundness levels. Practical proof systems therefore
often perform bulk arithmetic over a small base field while drawing
soundness-critical challenges from a larger extension field. For example,
Plonky3 supports Goldilocks and corresponding extension fields, while RISC Zero
uses BabyBear with a degree-4 extension~\cite{plonky3,risczero}. This design
keeps most prover arithmetic narrow while providing a substantially larger
challenge space for soundness-critical checks.
\tool{}'s small-width rows characterize that regime, isolating the
effect of limb count on GPU behavior while holding the protocol, the workload,
and the correctness criterion fixed. They use the same generic Montgomery arithmetic as the 256-bit
configurations rather than exploiting the special-form reductions available
for the Goldilocks prime and $2^{127}-1$. The small-width results therefore vary only the
limb count within one arithmetic implementation, rather than modeling a
field-specialized small-field prover. The 256-bit rows use the BLS12-381 scalar
field and cover the classical setting, where the base field alone carries
soundness.

Two scoping consequences follow. First, \tool{} measures base-field SumCheck
only; extension-field challenge arithmetic requires mixed-width kernels and is
out of scope. Second, a field in \tool{} is a parameter set rather than a kernel
property, consisting of a modulus, a limb count, and a Montgomery constant
shared by both backends, so adding a field requires no kernel changes.
\section{Design and Implementation}
\label{sec:design}

\subsection{Workload Specifications and Static Kernels}
\label{sec:design:spec}

The core abstraction in \tool{} is a \emph{workload specification}: a
backend-independent description of a polynomial constraint. Each specification
assigns a stable identifier, records the polynomial expression, and lowers it
into term metadata describing which variables appear in each term, how terms
are grouped, and which coefficients apply. Both backends consume the same
metadata, so a named workload denotes the same computation on either path.

\tool{} performs no dynamic expression generation in the measured path. Each
identifier maps to matched static implementations, which keeps three concerns
separate that are easy to conflate: the round count $r$ fixes the initial table
size $N = 2^r$; the field width fixes the number of limbs per element; and the
workload fixes which input tables, coefficients, and product chains are
evaluated per entry. Static kernels make results reproducible, make generated code
inspectable, and ensure the benchmark measures kernel execution rather than a
runtime expression compiler.

The suite has two tiers. The \emph{diagnostic} tier isolates individual
expression patterns: a single product, a length-three product chain, repeated
input symbols, and independent product terms summed together. It calibrates the
infrastructure and attributes behavior to one structural feature at a time. The
\emph{zkPHIRE} tier is the primary workload set: 25 constraints adapted from a
published SumCheck accelerator~\cite{zkphire2025}, spanning Spartan-style
satisfiability, Halo2 curve and point-addition gates, HyperPlonk zero- and
permutation-checks, their Jellyfish variants, and a polynomial opening check.
The tier is a spread across four structural axes that
independently stress a GPU: the number of distinct input tables (2 to 22), the
polynomial degree (2 to 7), the length of the longest product chain, and the
count of simultaneously live intermediates.

\subsection{Kernel Optimizations}
\label{sec:design:opts}

Both backends implement the same SumCheck round: the active table is paired into
even and odd entries, the round polynomial is evaluated at its $D+1$
interpolation points and accumulated, and the table is folded using the round
challenge. Holding this structure fixed isolates differences in host-side
orchestration and generated kernels rather than differences in the SumCheck
algorithm. Within this common structure, \tool{} exposes the optimizations below
as compile-time controls, each with a matched baseline and optimized path. No
runtime dispatch remains in the measured kernels, allowing each optimization to
be evaluated independently on identical workloads.

\subsubsection{\textbf{Specialized Point Interpolation}}
\label{sec:design:interp}

Interpolating input tables at the round's evaluation points is the dominant
per-element cost. Applying Equation~\ref{eq:interp} directly requires a modular
subtraction, a Montgomery multiplication, and a modular addition per point per
input table, so a workload referencing \texttt{NUM\_INPUTS} tables at degree
$D$ performs
\begin{equation}
  \label{eq:naive-cost}
  \texttt{NUM\_INPUTS} \times (D+1)
\end{equation}
Montgomery multiplications per (even, odd) pair for interpolation alone. At 256 bits each expands
into 64 limb multiply-accumulates, carry propagation, and a reduction pass.

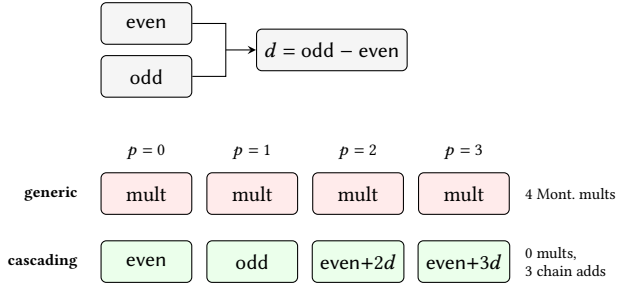
\begin{figure}[t]
  \centering
  \begin{tikzpicture}[
    font=\small,
    box/.style={draw, rounded corners=2pt, minimum height=5.5mm,
                minimum width=12mm, inner sep=3pt},
    val/.style={box, fill=black!4},
    gen/.style={box, fill=red!8},
    spec/.style={box, fill=green!8},
  ]
    \node[val] (e) at (0,0.35)  {$\mathsf{even}$};
    \node[val] (o) at (0,-0.35) {$\mathsf{odd}$};
    \coordinate (j) at (1.05,0);
    \draw (e.east) -| (j);
    \draw (o.east) -| (j);
    \node[val, anchor=west] (d) at (1.45,0) {$d=\mathsf{odd}-\mathsf{even}$};
    \draw[-stealth] (j) -- (d.west);

    \foreach \p/\x in {0/0, 1/1.4, 2/2.8, 3/4.2} {
      \node[font=\scriptsize] at (\x,-1.35) {$p=\p$};
    }

    \node[anchor=east, font=\scriptsize\bfseries] at (-0.8,-1.9) {generic};
    \foreach \x in {0, 1.4, 2.8, 4.2} { \node[gen] at (\x,-1.9) {mult}; }
    \node[anchor=west, font=\scriptsize] at (4.9,-1.9) {4 Mont.\ mults};

    \node[anchor=east, font=\scriptsize\bfseries] at (-0.8,-2.8) {cascading};
    \node[spec] at (0,-2.8)   {$\mathsf{even}$};
    \node[spec] at (1.4,-2.8) {$\mathsf{odd}$};
    \node[spec] at (2.8,-2.8) {$\mathsf{even}{+}2d$};
    \node[spec] at (4.2,-2.8) {$\mathsf{even}{+}3d$};
    \node[anchor=west, font=\scriptsize, align=left] at (4.9,-2.8) {0 mults,\\3 chain adds};
  \end{tikzpicture}
  \caption{Interpolating one input table at the four points of a degree-3
  round. The difference $d$ is formed once. The generic path then issues a
  Montgomery multiplication per point, whereas successive points differ by
  exactly $d$, so the specialized path returns $\mathsf{even}$ and
  $\mathsf{odd}$ directly and reaches the rest by modular addition.}
  \label{fig:interp}
\end{figure}

Substituting a compile-time point into Equation~\ref{eq:interp} exposes the
structure. Writing $d = \mathsf{odd} - \mathsf{even}$, the equation becomes
$\mathsf{even} + p \cdot d$, so $p=0$ returns $\mathsf{even}$, $p=1$ returns
$\mathsf{odd}$, and each subsequent point exceeds the previous one by exactly
$d$. For a point $p\geq2$, the multiple $p d$ can therefore be formed from $d$ with
$p-1$ modular additions, avoiding multiplication altogether.
Figure~\ref{fig:interp} shows this for a degree-3 round. \tool{} takes the
additive path for $2 \le p \le 15$ and the generic Montgomery path beyond,
where the chain grows long enough that a multiplication becomes competitive;
with a maximum degree of 7 in the suite, the fallback is never taken.

Summing over the points for one (even, odd) pair of one input table replaces $D+1$ Montgomery
multiplications with
\begin{equation}
  \label{eq:cascade-cost}
  \sum_{p=2}^{D} (p-1) \;=\; \frac{D(D-1)}{2}
\end{equation}
modular additions. For $p\geq2$, both implementations compute the subtraction
forming $d$ and the final addition of $\mathsf{even}$; the specialized cases
$p=0$ and $p=1$ omit these operations as well.
Table~\ref{tab:interp-cost} instantiates this for the degrees present in the
suite. At $D=7$, the transformation removes all 8 Montgomery multiplications per
input table in exchange for 21 modular additions, and on the widest workload,
which references 22 tables, it removes 176 multiplications per (even, odd) pair per
round.

\begin{table}[t]
  \centering
  \caption{Interpolation cost per input table per round. Cascading addition
  removes every Montgomery multiplication at the degrees present in the suite,
  trading them for modular additions. Counts include the cascading additions
  used to form $pd$; for $p\ge2$, the subtraction forming $d$ and final
  addition of $\mathsf{even}$ are common to both implementations.}
  \label{tab:interp-cost}
  \begin{tabular}{@{}rcc@{}}
    \toprule
    Degree $D$ & Mont.\ mults removed & Cascading mod.\ adds \\
    \midrule
    1 & 2 & 0  \\
    2 & 3 & 1  \\
    3 & 4 & 3  \\
    5 & 6 & 10 \\
    7 & 8 & 21 \\
    \bottomrule
  \end{tabular}
\end{table}

The optimization is a compile-time specialization in both backends. The point
mode and point index are resolved before code generation, so the emitted kernel
for $p=0$ contains only the direct return, the kernel for $p=1$ returns the odd
value, and kernels for $2 \le p \le 15$ contain the additive chain and no multiplication. No
branch survives into the measured code. The generic path remains selectable
through \texttt{-{}-point-mode} for comparison.

\subsubsection{\textbf{Limb-Major Table Layout}}
\label{sec:design:layout}

\begin{figure}[t]
  \centering
  \begin{tikzpicture}[font=\scriptsize,
    cell/.style={draw, minimum width=2.4mm, minimum height=2.4mm, inner sep=0pt},
    hot/.style={cell, fill=blue!25},
    cold/.style={cell, fill=black!3},
  ]
    \node[anchor=west, font=\small\bfseries] at (0,0.95) {element-major};
    \foreach \i in {0,...,31} {
      \pgfmathtruncatemacro{\ishot}{mod(\i,8)==0 ? 1 : 0}
      \ifnum\ishot=1 \node[hot] at (\i*2.4mm,0) {};
      \else          \node[cold] at (\i*2.4mm,0) {}; \fi
    }
    \foreach \e/\x in {0/0, 1/8, 2/16, 3/24} {
      \draw[decorate, decoration={brace, amplitude=2pt}]
        ({\x*2.4mm-1.2mm},2.0mm) -- ({(\x+8)*2.4mm-1.2mm},2.0mm)
        node[midway, above, yshift=-1pt, font=\tiny] {elem.\ \e};
    }

    \node[anchor=west, font=\small\bfseries] at (0,-1.15) {limb-major};
    \foreach \i in {0,...,31} {
      \ifnum\i<4 \node[hot] at (\i*2.4mm,-1.75) {};
      \else      \node[cold] at (\i*2.4mm,-1.75) {}; \fi
    }
    \draw[decorate, decoration={brace, amplitude=2pt, mirror}]
      (-1.2mm,-19.9mm) -- (8.4mm,-19.9mm)
         node[midway, below, yshift=-2pt, font=\tiny] {limb 0 of elem.\ $0\ldots3$};
  \end{tikzpicture}
  \caption{The same four words under both layouts at $L=8$: limb~0 of four
  consecutive field elements, as read by four consecutive lanes. Element-major
  scatters them at stride $L$; limb-major makes them contiguous. The pattern
  extends to a full 32-lane warp.}
  \label{fig:layout}
\end{figure}
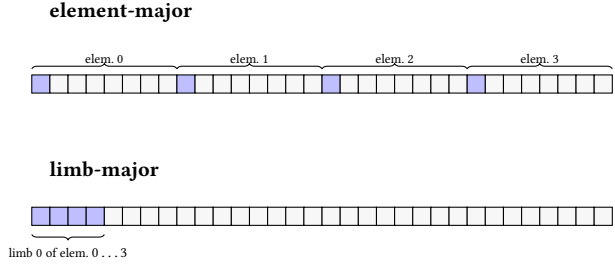

A table of $N$ field elements of $L$ limbs each can be stored two ways. The
default \emph{element-major} layout places all limbs of one element
contiguously,
\begin{equation}
  \texttt{addr} = \texttt{base} + \texttt{idx} \cdot L + \texttt{limb},
\end{equation}
which is natural for scalar arithmetic but places neighbouring lanes $L$ words
apart; at $L=8$ that is a 32-byte stride. The \emph{limb-major} layout stores
each limb position as a contiguous plane,
\begin{equation}
  \texttt{addr} = \texttt{base} + \texttt{limb} \cdot \texttt{row\_stride}
                  + \texttt{idx},
\end{equation}
so a single lane gathers one element's limbs from $L$ planes, but a warp
reading the same limb position across 32 consecutive elements touches one
contiguous 128-byte region, as shown in Figure~\ref{fig:layout}. Both backends
select the layout at compile time, so the emitted kernel contains only one
address computation and no per-access dispatch. The choice is exposed through
\texttt{-{}-layout}. Section~\ref{sec:eval:width} measures its effect.

\subsection{Backend Realization}
\label{sec:design:backends}

The two backends express the same static specialization in their natural
idioms. In Triton, the workload identifier, limb count, layout mode, and point
mode are \texttt{constexpr} parameters, while round-dependent quantities such
as the active table size are passed as runtime values. This distinction is
necessary because SumCheck shrinks its table every round, and
treating the active size as a specialization key would make the benchmark
measure recompilation instead of steady-state execution. Benchmark scripts
expose \texttt{TRITON\_CACHE\_DIR} so cold and warm behavior are recorded
separately.

In CUDA, field width is selected through template cases for one, two, four, and
eight limbs, and workload identifiers dispatch to matched static evaluators.
The CUDA path builds as native executables rather than Python bindings; the
benchmark driver launches them, parses their CSV output, and joins rows with
the Triton results by workload, bit width, and round count.

Our CUDA implementation is intended to be a strong compiled baseline rather
than a hand-written PTX kernel. We specialize each evaluation kernel to the
workload, table count, field width, layout, and interpolation point, which lets
the compiler fully unroll the limb loops. Field multiplication uses CIOS
Montgomery reduction with 64-bit accumulation, while addition and subtraction
use explicit carry chains followed by a conditional correction. The code is
compiled with \texttt{nvcc -O3} for the native architecture and otherwise leaves
register allocation and instruction selection to the compiler. We swept block
sizes from 64 to 512 threads and use 128 throughout, which is either the fastest
choice or within 1\% of fastest for the evaluation kernels. The CUDA and Triton paths
therefore perform the same underlying field computation, while retaining three
implementation differences: the block reduction
(Section~\ref{sec:eval:registers}), modular correction, and the C++ versus
Python host loops.
\section{Methodology}
\label{sec:method}

\subsection{Correctness Gating}

We validate correctness in two complementary ways. Both implementations are
first checked against a CPU reference that evaluates each workload polynomial
independently using Python big integers. This covers every reported workload
and field width at $r=3$ and $r=10$, with fixed and SHA3 challenges; all 381
cases per backend match.

\begin{figure}[t]
  \centering
  \begin{tikzpicture}[font=\scriptsize,
    b/.style={draw, rounded corners=2pt, align=center, inner sep=3pt,
              minimum height=6mm},
    spec/.style={b, fill=black!5, minimum width=52mm},
    be/.style={b, fill=red!6, minimum width=24mm},
    ref/.style={b, fill=blue!6, minimum width=34mm},
    gate/.style={b, fill=green!8, minimum width=52mm},
  ]
    \node[spec] (s) at (0,0)
      {workload spec: id, expression, term metadata};

    \node[be] (t) at (-1.6,-1.2) {Triton kernel\\[-1pt]
      \tiny \texttt{constexpr}: id, $L$, layout, point};
    \node[be] (c) at ( 1.6,-1.2) {CUDA executable\\[-1pt]
      \tiny templates $L\!\in\!\{1,2,4,8\}$};

    \draw[-stealth] (s.south) -- ++(0,-0.2) -| (t.north);
    \draw[-stealth] (s.south) -- ++(0,-0.2) -| (c.north);

    \node[ref, minimum width=22mm] (r) at (0,-2.2)
      {CPU big-integer reference\\[-1pt]
       \tiny validates both backends};
    \draw[dashed] (r.north west) -- (t.south east);
    \draw[dashed] (r.north east) -- (c.south west);

    \node[gate] (g) at (0,-3.3)
      {join on (workload, width, rounds);\\[-1pt]
       emit iff Triton and CUDA checksums agree};
    \draw[-stealth] (t.south) -- (t.south |- g.north);
    \draw[-stealth] (c.south) -- (c.south |- g.north);

    \node[b, fill=black!3, minimum width=34mm] (k) at (0,-4.4)
      {row, or skip with reason};
    \draw[-stealth] (g.south) -- (k.north);
  \end{tikzpicture}
  \caption{Correctness gating in the benchmark harness. Reported
  configurations require matching CUDA and Triton checksums, while both
  implementations are independently validated against the CPU big-integer
  reference.}
  \label{fig:harness}
\end{figure}
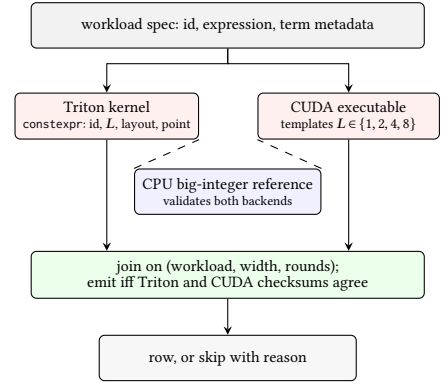

The performance sweep then carries a checksum over the round-polynomial
evaluations produced by each backend. CUDA and Triton results are joined by
workload, field width, and round count, and a configuration enters the reported
results only when the two checksums agree.
Configurations that fail, exceed the 30-minute limit, or exhaust device memory
are retained with their status and failure reason, and 
Section~\ref{sec:eval} accounts for these cases in the reported sweep. Figure~\ref{fig:harness} summarizes
this flow. Together, these checks guard against errors in carry propagation,
reduction, layout indexing, and challenge generation that could otherwise leave
timing results looking entirely plausible while changing the underlying
computation.

\subsection{Timing}

Unless otherwise stated, all results use specialized interpolation points, the
element-major layout, and whole-run timing. \emph{Whole-run} timing covers one
complete SumCheck execution: each round launches one evaluation kernel per
interpolation point, reduces the resulting partial sums across multiple kernels,
synchronously copies each point result to the host, and folds the table for the
next round. Table generation and Montgomery encoding occur outside the timed
region.

Each configuration executes once outside the measurement set to absorb Triton
JIT compilation and CUDA module loading, followed by two warm-up runs and ten
timed runs. For the H100 sweeps, both backends use synchronized host timers,
\texttt{time.perf\_counter} for Triton and
\texttt{std::chrono::steady\_clock} for CUDA, and we report the median of the
ten measurements. CUDA kernels use 128 threads per block, while Triton programs
use 128 lanes with the default four warps; corresponding kernels use matched
grid dimensions. Each configuration is limited to 30 minutes.

For the GPU-kernel measurements in Section~\ref{sec:eval:registers}, Nsight
Systems records the duration of every kernel. We sum kernel durations within
each SumCheck execution, average this total over the 13 executions of a
configuration, and report medians across constraints. Per-configuration kernel
times are stable to within 1\%, whereas wall-clock time is more sensitive to
host state; all wall-clock ratios compared within a table therefore come from
the same measurement session.

\subsection{Platform}

Our primary evaluation platform is an NVIDIA H100 PCIe with 80\,GB of memory,
114 SMs, and compute capability 9.0, running CUDA 12.9, PyTorch 2.12.0, and
Triton 3.7.0. We use the H100 for the main performance results and mechanism
studies in Section~\ref{sec:eval}. To assess portability across GPU
generations, we also use two independent A100 systems: an A100-SXM4-80GB
platform for a full performance sweep and an A100 PCIe system for GPU-kernel
measurements. Because host-inclusive timing depends on the surrounding system,
wall-clock comparisons are kept within each machine.
Table~\ref{tab:machine-specs} in Appendix~\ref{app:platform} records the
complete hardware and software environments, and the experiment harness captures the associated metadata,
logs, and result files for each run.

\section{Evaluation}
\label{sec:eval}

We evaluate 800 configurations on the H100, spanning 25 zkPHIRE-derived
constraints, four arithmetic widths, and round counts
$r \in \{14,16,\ldots,28\}$ ($r$ also indicates the starting size of each table, $2^r$ field elements). Of these, 758 complete with matching Triton and
CUDA checksums and form the results below. The remaining configurations are
explained by either memory capacity or compilation cost. Thirty-six exceed the
80\,GB device capacity: each run must hold an input table together with two
full-size fold buffers, for a footprint of
$3 \times \mathit{vars} \times 2^{r} \times L \times 4$ bytes. The largest
configuration that completes uses 77.3\,GB, while the skipped cases would
require 90--567\,GB. The final six configurations are the 256-bit instances of
\texttt{zk\_jellyfish\_zerocheck\_hp} at $r=14$--$24$, where Triton compilation
exceeds the 30-minute limit; Section~\ref{sec:eval:compile} examines this
compiler-scaling limit in more detail. 
The results support four observations. CUDA is faster than Triton on the
matched runs overall, but the size of the gap depends strongly on expression
structure and arithmetic width; at the widest fields and largest tables the two
backends reach parity and the ordering sometimes reverses; the gap comes
mostly from host launch overhead and a Triton reduction idiom, with register
pressure a secondary factor; and the zkPHIRE ASIC
provides context for the remaining distance to domain-specialized hardware.

\subsection{Triton versus CUDA}
\label{sec:eval:diag}

\begin{figure}[t]
  \centering
  \includegraphics[width=\columnwidth]{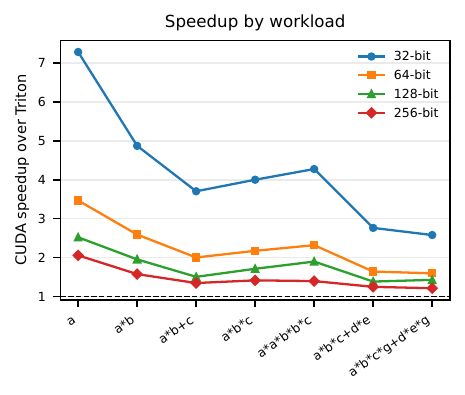}
  \caption{CUDA speedup over Triton on the diagnostic tier (H100). The gap
  narrows with arithmetic width across series and generally with expression
  complexity from left to right.}
  \label{fig:diag}
\end{figure}

Figure~\ref{fig:diag} compares the backends across the diagnostic tier, whose
polynomials range from a single table read to a seven-symbol expression with
two product chains. CUDA remains faster throughout, while the overall gap
generally narrows as arithmetic intensity and expression complexity increase:
at 32 bits it falls from $7.3\times$ on a single symbol to $2.6\times$ on the
longest chain. 

\subsection{Field Width and Representation}
\label{sec:eval:width}

Table~\ref{tab:by-width} aggregates the 758 validated zkPHIRE configurations by
arithmetic width, and both summary statistics show the same progression as the
field widens. CUDA's median advantage falls from $3.50\times$ at 32 bits to
$2.93\times$, $2.17\times$, and $1.37\times$ at 64, 128, and 256 bits,
respectively; over the same widths, its narrowest margin falls from
$1.48\times$ to $1.11\times$, $0.86\times$, and $0.65\times$, crossing
parity for the first time at 128 bits.

The two widest fields also contain a growing set of configurations where the
backends are effectively comparable. At 256 bits, 30 of 171 configurations
(18\%) finish within $1.1\times$ of CUDA and 19 favor Triton, while at 128 bits
the corresponding counts are 11 of 189 and 7; by contrast, no 32- or 64-bit
configuration comes within $1.1\times$, and CUDA leads every one of them by at
least $1.109\times$. These wins are generally larger than run-to-run variation:
19 of the 26 favor Triton by more than 5\%, with the largest advantages
appearing for \texttt{zk\_complete\_add\_3} at $r=26$ ($0.650\times$) and
$r=24$ ($0.681\times$), followed by \texttt{zk\_complete\_add\_6} at $r=26$
($0.739\times$).

The near-parity configurations also follow a clear workload and table-size
pattern rather than appearing uniformly across the sweep. Most wins come from
the addition-constraint family, \texttt{zk\_complete\_add\_1} through
\texttt{\_6} and \texttt{zk\_incomplete\_add\_2}, with two additional 128-bit
wins from \texttt{zk\_jellyfish\_zerocheck\_hp}; the broader near-parity set
also includes the HyperPlonk vanilla and Jellyfish checks. At 256 bits,
near-parity first appears at $r=18$ and Triton wins begin at $r=20$, increasing
to seven wins at $r=26$, while at 128 bits the corresponding thresholds move to
$r=22$ and $r=24$. No 256-bit win appears at $r=28$, although only four of the
25 constraints still fit in memory at that size, so the overall pattern is one
of convergence at wider arithmetic and larger tables, most clearly among the
suite's more arithmetic-intensive constraints.

We evaluate both kernel optimizations on H100 across five constraints at 32
and 256 bits, with $r=20$ and $24$. Specialized interpolation improves 256-bit
evaluation-kernel performance by $1.65$--$1.75\times$ in CUDA and
$1.55$--$1.59\times$ in Triton, compared with only $1.04$--$1.06\times$ at 32
bits, where single-limb multiplication is relatively inexpensive. With
specialized interpolation enabled, limb-major storage provides a further
1--5\% improvement at 256 bits; under generic interpolation, its effect is
small.

\begin{table}[t]
  \centering
  \caption{The 758 validated zkPHIRE configurations on the H100 by arithmetic width. Best
  ratio is CUDA's narrowest margin at that width; values below $1.0$ mean
  Triton is faster.}
  \label{tab:by-width}
  \begin{tabular}{@{}rrrrr@{}}
    \toprule
    Width & Median & Best ratio & Wins & Within $1.1\times$ \\
    \midrule
    32  & 3.50$\times$ & 1.481 & 0  & 0  \\
    64  & 2.93$\times$ & 1.109 & 0  & 0  \\
    128 & 2.17$\times$ & 0.855 & 7  & 11 \\
    256 & 1.37$\times$ & 0.650 & 19 & 30 \\
    \midrule
    All & ---          & 0.650 & 26 & 41 \\
    \bottomrule
  \end{tabular}
\end{table}
\subsection{Decomposing the CUDA--Triton Gap}
\label{sec:eval:registers}

To understand where the end-to-end gap arises, we use Nsight Systems to
separate GPU execution from the host-side portion of each run.
Table~\ref{tab:split} summarizes all 25 constraints at 32 bits and 24 at 256
bits. CUDA and Triton launch the same sequence of 313--1303 kernels per
SumCheck run, while the host cost of each launch differs substantially:
Triton's Python driver incurs 11--22\,$\mu$s per launch, compared with
3--4\,$\mu$s for the CUDA executable. This overhead dominates the
smaller-field regime. At 32 bits and $r=20$, 91\% of Triton's wall-clock time
contains no GPU-kernel execution, reducing CUDA's $4.0\times$ end-to-end
advantage to $1.38\times$ when comparing kernel time alone.

Most of the remaining kernel-level difference comes from the block reduction.
Each evaluation block contains 128 threads, producing 128 partial sums that
ultimately have to be combined into one field element. In our original Triton
implementation, these partial sums are materialized so that every thread retains
the complete set and independently carries out the reduction, turning the 127
modular additions needed once per block into 127 additions per thread. 
Replacing the reduction with \texttt{tl.reduce} lets Triton express the
block-wide reduction directly; every checksum remains unchanged, while
Triton's median kernel time falls to $0.92$--$1.05\times$ of CUDA across the
configurations in Table~\ref{tab:split}.

Register allocation contributes a smaller residual effect. The measured
kernels do not spill, although at 32 bits Triton uses 44--46 registers per
thread compared with CUDA's 18--32. Despite this difference, all 25
constraints have the same CUDA/Triton occupancy ratio while their
evaluation-kernel gaps span $1.4\times$ to $5.9\times$. Constraining CUDA to
Triton's occupancy by padding shared memory slows it by at most $1.25\times$
across four 32-bit constraints, which bounds the performance attributable to
occupancy in this regime. At 256 bits, the effect becomes more pronounced for
some kernels and accounts for nearly the full residual gap in
\texttt{zk\_vanilla\_permcheck\_hp}.

Across the profiled evaluation kernels, DRAM throughput reaches only
6--42\% of peak, ruling out off-chip bandwidth as the source of the observed
gap.

The relative importance of these costs changes as field arithmetic widens.
Launch overhead remains roughly fixed per kernel, reduction cost grows
approximately linearly with the number of limbs, and multi-limb Montgomery
multiplication grows quadratically. Consequently, arithmetic increasingly
dominates both host overhead and reduction cost: at 256 bits and $r=24$, the
fraction of Triton's wall-clock time without GPU-kernel execution falls to
10\%, while CUDA's kernel-time advantage narrows to $1.12\times$ and to
$1.01\times$ after replacing the reduction with \texttt{tl.reduce}.

\begin{table}[t]
  \centering
  \caption{CUDA--Triton runtime breakdown on H100. Entries are Triton/CUDA
  ratios; values below $1\times$ favor Triton. ``No kernel'' is Triton's
  wall-clock fraction with no GPU kernel executing.}
  \label{tab:split}
  \small
  \begin{tabular}{@{}lrrrr@{}}
    \toprule
    Width, $r$ & Wall T/C & Kernel T/C & +\texttt{tl.reduce} T/C & No kernel \\
    \midrule
    32, 20  & 4.02 & 1.38 & 0.93 & 91\% \\
    32, 24  & 3.17 & 2.20 & 1.05 & 60\% \\
    256, 20 & 1.52 & 1.03 & 0.92 & 49\% \\
    256, 24 & 1.24 & 1.12 & 1.01 & 10\% \\
    \bottomrule
  \end{tabular}
\end{table}

\subsection{A Compile-Time Limit in Triton}
\label{sec:eval:compile}

\texttt{zk\_jellyfish\_zerocheck\_hp}, which references 22 input tables at
degree 7, completes on both backends from 32 through 128 bits. At 256 bits,
CUDA still compiles the specialized-point kernels to 226--236 registers per
thread without spilling, whereas Triton compilation grows from seconds at
32 bits to minutes at 64--128 bits and hours at 256 bits. At 64 and 128 bits,
96--99\% of this time is spent in \texttt{TritonGPUCoalesce}\footnote{\url{https://github.com/triton-lang/triton/blob/v3.7.0/lib/Dialect/TritonGPU/Transforms/Coalesce.cpp\#L71}; the per-operation slice is built in \href{https://github.com/triton-lang/triton/blob/v3.7.0/lib/Dialect/TritonGPU/Transforms/CoalesceUtils.cpp\#L45}{\texttt{CoalesceUtils.cpp}, line~45}.}. The pass
repeatedly constructs def-use slices for loads and stores, causing its cost to
rise rapidly as the fully unrolled kernel expands. We observe the same
behavior in the latest Triton releases, indicating a compiler scalability
limit for large unrolled kernels rather than a regression specific to one
version.

\subsection{zkPHIRE-Derived Benchmark Results}
\label{sec:eval:family}

At 256 bits and $r=24$ (Table~\ref{tab:family}), CUDA retains a $1.15\times$
geometric-mean advantage over Triton across the 24 constraints matched to
zkPHIRE, although the gap varies substantially by workload family, from
$1.50\times$ for the Verifiable ASICs/Spartan constraints to $1.03\times$ for
complete addition. The addition constraints remain the strongest case for
Triton, with the two backends within three percent on complete addition, but
the convergence is no longer confined to that family: addition constraints
account for 34 of the 41 configurations within $1.1\times$ and 24 of the 26
Triton wins, while several HyperPlonk configurations also approach parity.
Relative to zkPHIRE's reported ASIC results~\cite{zkphire2025}, CUDA remains
$4.60\times$ slower in geometric-mean runtime ($74.16$ versus $16.11$\,ms),
leaving substantial headroom beyond what either GPU implementation recovers.

\begin{table}[t]
  \centering
    \caption{zkPHIRE-derived constraints at 256 bits, $r=24$. Geometric-mean
  runtimes in milliseconds; ASIC is the zkPHIRE
  accelerator~\cite{zkphire2025}. Jellyfish zerocheck does not compile within the time
  limit at eight limbs (Section~\ref{sec:eval:compile}), hence $n=1$.}
  \label{tab:family}
  \begin{tabular}{@{}lrrrrr@{}}
    \toprule
    Family & $n$ & Triton & CUDA & Speedup & ASIC \\
    \midrule
    Verifiable / Spartan  & 3  & 25.60  & 17.11  & 1.50$\times$ & 6.31  \\
    Halo2 curve checks    & 3  & 56.21  & 41.43  & 1.36$\times$ & 8.86  \\
    Incomplete addition   & 2  & 100.85 & 92.49  & 1.09$\times$ & 23.06 \\
    Complete addition     & 12 & 104.60 & 101.44 & 1.03$\times$ & 19.95 \\
    HyperPlonk vanilla    & 2  & 129.90 & 113.84 & 1.14$\times$ & 19.81 \\
    HyperPlonk Jellyfish  & 1  & 406.30 & 388.52 & 1.05$\times$ & 30.09 \\
    Opening check         & 1  & 56.10  & 42.13  & 1.33$\times$ & 21.29 \\
    \midrule
    All matched           & 24 & 84.96  & 74.16  & 1.15$\times$ & 16.11 \\
    \bottomrule
  \end{tabular}
\end{table}

\subsection{Portability to the A100}
\label{sec:eval:a100}

The A100-SXM4 sweep shows the same width-dependent convergence as the H100:
CUDA's median advantage falls from $4.39\times$ at 32 bits to $3.58\times$,
$2.58\times$, and $1.60\times$ as the field widens, while 26 of 165 validated
256-bit configurations finish within $1.1\times$ of CUDA, including 17 Triton
wins; 16 of these are also wins on the H100. The kernel-level decomposition
carries across generations as well. On an A100 PCIe, CUDA's kernel-time
advantage is $1.42\times$ at $r=20$ and $2.37\times$ at $r=24$ for 32-bit
fields, while \texttt{tl.reduce} brings Triton's median kernel time to
$0.97$--$1.08\times$ of CUDA's.

\section{Related Work}
\label{sec:related}

\paragraph{GPU acceleration of proof systems.}
Prior GPU work has primarily targeted MSMs and NTTs, the kernels that dominate pairing-based provers~\cite{cuzk,gzkp}. Production libraries such as ICICLE~\cite{icicle} package these and other proving primitives as reusable GPU backends across several curves and field sizes. The optimizations for MSM and NTT are largely kernel-specific: bucket-method scheduling and point accumulation for MSM, and butterfly scheduling, twiddle-factor organization, and memory-layout optimizations for NTT. These techniques do not directly transfer to SumCheck, whose dominant operation is instead a sequence of dependent table folds. Recent systems have also accelerated NTTs, Merkle trees, and other primitives in hash-based provers~\cite{megazk,batchzk}.

\paragraph{SumCheck on GPUs.}
Recent work has begun accelerating SumCheck on GPUs, but has considered a relatively narrow range of SumCheck polynomials. BatchZK~\cite{batchzk} implements a pipelined GPU SumCheck for a single multilinear polynomial, optimizing repeated table reductions and folds for high-throughput batch proving. Chimera~\cite{chimera} accelerates SumCheck within Spartan, keeping the active polynomial state resident on the GPU across profitable rounds and falling back to the CPU for small domains; consequently, its evaluation is restricted to Spartan's low-degree R1CS SumChecks, whose polynomials have degree at most three. ICICLE~\cite{icicle} provides a more general CUDA SumCheck primitive with a programmable combine function, but limits the number of constituent multilinear polynomials and does not characterize performance across high-degree SumChecks. These works demonstrate that SumCheck can benefit substantially from GPU acceleration, but study it within a fixed library or prover configuration. \tool{} instead isolates SumCheck as a standalone kernel, allowing protocol structure, field representation, polynomial degree, and implementation strategy to be varied independently.



\paragraph{Characterization and primitive comparison.}
ZKProphet~\cite{zkprophet} presents a performance study of ZKPs on GPUs,
finding that prover arithmetic executes almost entirely on the 32-bit integer
pipeline and is limited by data dependencies and available integer compute; it
covers MSM and NTT. 
Other work provides a 
CPU microarchitectural characterization of Groth16, profiling pipeline stalls,
memory behavior, instruction mix, and multicore scalability across several CPUs
and elliptic curves \cite{cpu_zkp_perf}. 
Complementary work compares NTT- and SumCheck-based
primitives under matched on-chip SRAM and off-chip bandwidth budgets and finds
no universal winner, with the better primitive depending on polynomial degree
and available memory~\cite{whenproofs}. Both hold the implementation fixed and
vary the kernel or the budget; \tool{} holds the protocol and workload fixed and
varies the programming model and arithmetic width on real hardware.

\paragraph{Specialized hardware.}
A growing body of work has focused on ASIC and commodity acceleration of kernels present in ZKPs, and has focused primarily on MSMs and NTTs ~\cite{priormsm, distmsm, legozk, unizk, pipezk, gypso, MSMAC, tches_ntt_msm, zhang2025code, zhang2025micro, elastic_msm, sam, nocap, vroom, szkp, locus}, with other work in cryptographic computing focused on fully homomorphic encryption and multi-party computation \cite{kim2022_BTS, sharp, rpu, haac, karthik, orion, ciflow, mo2023towards, ark, jha2024deepreshape, osiris, anaheim}.

Recent work proposed domain-specific accelerators for
SumCheck-based provers~\cite{zkspeed2025,zkphire2025}. \tool{}'s primary workload tier derives from
the zkPHIRE constraint set, and Section~\ref{sec:eval:family} uses its
published numbers as context for the distance between commodity GPUs and specialized silicon. Existing ZKP benchmarking efforts compare complete proof system implementations at the application level; \tool{} operates one level down, at the granularity of a single prover kernel with matched backends and correctness gating.
\section{Conclusion}
\label{sec:conclusion}

The cost of programming-model abstraction in SumCheck is not fixed. Across 758
validated H100 configurations, CUDA's median advantage over Triton falls from
$3.50\times$ at 32 bits to $1.37\times$ at 256 bits; 18\% of the 256-bit
configurations finish within $1.1\times$ of CUDA, compared with none at 32 or
64 bits, and an A100 shows the same width-dependent convergence. Much of the
difference comes from host launch overhead and Triton's original block
reduction, fixed costs that are increasingly amortized as multi-limb Montgomery
arithmetic grows; replacing the reduction with \texttt{tl.reduce} brings
Triton's median kernel time within 8\% of CUDA on both GPUs.

For prover implementations at cryptographic widths, these results make a
Python-embedded DSL a practical choice, and arithmetic specialization benefits
both programming models. The abstraction does encounter a compiler-scaling
limit on the suite's widest constraint: CUDA compiles and runs the 256-bit,
22-table degree-7 kernel without spilling, while Triton requires hours per
kernel as \texttt{TritonGPUCoalesce} scales poorly with the fully unrolled
program.

Several directions remain. Cross-vendor evaluation through Triton's AMD ROCm
backend~\cite{rocm} would test whether the same trends hold beyond NVIDIA GPUs,
while extension-field arithmetic would broaden the study to provers over small
base fields. At narrow widths, CUDA Graphs or persistent kernels could further
reduce the launch overhead exposed by our measurements.

\begin{acks}
This work was supported by NSF CAREER award
\#2340137, NSF CIRC GRAND \#2450539, and NSF NeTs
\#2504400, and generous support from DTCC, AMD,
Google, and NY State Center for Advanced Technology in Telecommunications (CATT). The views, opinions, and/or findings expressed are
those of the authors and do not necessarily reflect the views of sponsors.
\end{acks}

\balance
\bibliographystyle{ACM-Reference-Format}
\bibliography{refs}

\appendix
\section{Experimental Platform}
\label{app:platform}

\begin{table}[H]
\centering
\caption{Experimental platforms. The H100 is the primary platform; the
A100-SXM4 system ran the full sweep, timing CUDA with CUDA events; the A100 PCIe system was used only for
GPU-kernel measurements. All three run nvcc 12.9, PyTorch 2.12.0+cu130 (CUDA
runtime 13.0), and Triton 3.7.0.}
\label{tab:machine-specs}
\footnotesize
\setlength{\tabcolsep}{3pt}
\begin{tabular}{@{}llll@{}}
\toprule
 & H100 (primary) & A100 (full sweep) & A100 PCIe \\
\midrule
GPU & H100 PCIe & A100-SXM4-80GB & A100 80GB PCIe \\
Memory, SMs & 80\,GB, 114 & 80\,GB, 108 & 80\,GB, 108 \\
Compute cap. & 9.0 & 8.0 & 8.0 \\
Driver & 570.148.08$^{*}$ & 580.126.20 & 565.57.01$^{*}$ \\
CPU & Xeon Plat.\ 8480+ & Xeon @ 2.20\,GHz & Xeon Gold 6338 \\
Cores, memory & 26, 237\,GB & 12, 167\,GB & 12, 127\,GB \\
OS & Ubuntu 22.04.5 & Ubuntu 24.04.4 & Ubuntu 22.04.5 \\
\bottomrule
\multicolumn{4}{@{}l}{\scriptsize $^{*}$Kernel driver; user-mode driver 580.126.20 (forward compatibility).}\\
\end{tabular}
\end{table}

\end{document}